\documentclass{optica-article}

\journal{opticajournal}

\articletype{Research Article}

\begin{document}

\title{Beyond the Rozanov bound on electromagnetic absorption via periodic temporal modulations}

\author{Zeki Hayran, and Francesco Monticone\authormark{*}}

\address{School of Electrical and Computer Engineering, Cornell University, Ithaca, New York 14853, USA}

\email{\authormark{*}francesco.monticone@cornell.edu}

\begin{abstract*} Incorporating time-varying elements into electromagnetic systems has shown to be a powerful approach to challenge well-established performance limits, for example bounds on absorption and impedance matching. So far, the majority of these studies have concentrated on time-switched systems, where the material undergoes instantaneous modulation in time while the input field is entirely contained within it. This approach, however, necessitates accurate timing of the switching event and limits how thin the system can ultimately be due to the spatial width of the impinging pulse. To address these challenges, here we investigate the \emph{periodic} temporal modulation of highly lossy materials, focusing on their relatively unexplored parametric absorption aspects. Our results reveal that, by appropriately selecting the modulation parameters, the absorption performance of a periodically modulated absorber can be greatly improved compared to its time-invariant counterpart, and can even exceed the theoretical bound for conventional electromagnetic absorbers, namely, the ``Rozanov bound''. Our findings thus demonstrate the potential of periodic temporal modulations to enable significant improvements in absorber performance while circumventing the limitations imposed by precise timing and material thickness in time-switched schemes, opening up new opportunities for the design and optimization of advanced electromagnetic absorber systems for various applications.
\end{abstract*}

\section{Introduction} \label{introduction}

Electromagnetic absorbers have become indispensable components in various applications across the electromagnetic spectrum, including energy harvesting, photovoltaics, sensing, radar cross-section reduction, electromagnetic interference shielding, and stealth technologies. Over the years, a wide range of engineered absorber structures have been developed to fulfill diverse operational requirements, including Salisbury screens \cite{salisbury1952absorbent}, Jaumann layers \cite{knott1980performance}, Dallenbach layers \cite{dallenbach1938reflection}, just to name a few examples from the applied electromagnetics literature. Moreover, recent advancements in nanotechnology and materials science have led to the emergence of novel absorber materials, such as graphene-based absorbers \cite{amin2013ultra}, carbon nanotube composites \cite{saib2006carbon}, and metal-dielectric metamaterials \cite{dayal2013design}. In addition, the integration of frequency selective surfaces \cite{sun2012broadband}, metasurfaces \cite{wakatsuchi2013waveform}, and photonic crystals \cite{nam2014solar} has further enhanced the performance of electromagnetic absorbers by offering improved impedance matching characteristics, and a higher degree of control with respect to incident polarization, frequency, angle, and other degrees of freedom. Numerous studies have also focused on the optimization of absorber structures through computational approaches such as genetic algorithms \cite{michielssen1993design}, particle swarm optimization \cite{cui2005application}, and deep learning techniques \cite{hou2020customized} to achieve superior performance and miniaturization. Furthermore, tunable absorbers that can adapt their absorption properties in real-time have also been investigated, such as liquid crystal-based  \cite{shrekenhamer2013liquid}, ferrite-based \cite{lei2016magnetically}, and phase-change materials-based \cite{kats2012ultra} absorbers. These materials have been shown to provide dynamic control of absorption spectra and enable reconfigurable absorption profiles to suit various applications. Despite the considerable progress in the development of electromagnetic absorbers, however, there still exist several challenges that require further investigation. Among these challenges are the demand for broadband absorption, efficient absorption at low frequencies, and thickness reduction without compromising the absorption performance \cite{zeng2020electromagnetic}.

\begin{figure*}
 \centering
  \includegraphics[ width=0.7\linewidth, keepaspectratio]{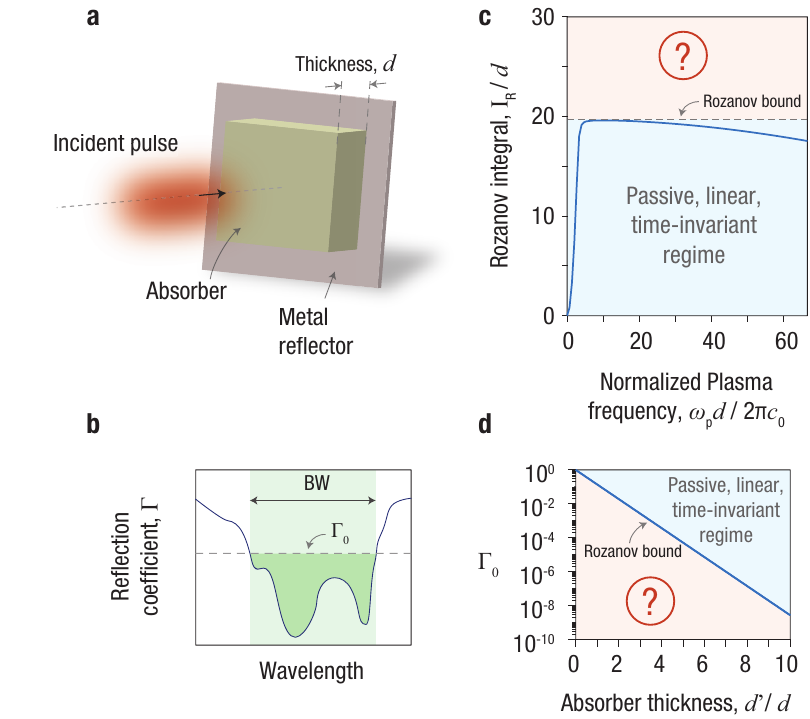}
\caption{(a)  Schematic representation of an electromagnetic pulse impinging on a non-magnetic absorber with thickness $d$ backed by a metallic mirror. (b) An illustrative reflection coefficient spectrum, with the absorption bandwidth dictated by the maximum tolerable reflection magnitude ($\Gamma_0$) from the absorber. (c) A practical example demonstrating the constraint on the integral $\mathrm{I}_\mathrm{R}$ imposed by the Rozanov bound for passive, linear, and time-invariant absorbers.  (d) Implications of the Rozanov bound for electromagnetic absorption, highlighting the typical necessity for increased absorber thickness to improve the absorption performance, for a given bandwidth. The normalized absorption bandwidth in this case is assumed to be 10 $d$. In (c) and (d), the considered material has a lossy Drude-type dispersion with a normalized damping coefficient equal to $\gamma d / 2\pi c_0 = 330$. In (d) the plasma frequency is equal to $\omega_{\mathrm{p}} d / 2\pi c_0 = 6$, where $c_0$ is the speed of light in vacuum.
}
    \label{fig1}
\end{figure*}

To understand the challenges associated with electromagnetic absorption, consider a non-magnetic absorbing slab characterized by thickness $d$ and complex permittivity $\epsilon(\lambda)$ (where $\lambda$ is the free space wavelength of the impinging wave), backed by a perfectly reflecting metal mirror, as depicted in Fig. \ref{fig1}(a). To determine the absorption bandwidth (BW) of this configuration, the reflection coefficient spectrum, $\Gamma(\lambda)$, can be employed by considering the maximum allowable reflection magnitude, $\Gamma_0$, as illustrated in Fig. \ref{fig1}(b). Although increasing the absorber thickness $d$ can generally extend BW, as in the case of long adiabatically tapered absorbers (e.g., standard anechoic panels), many practical applications call for a broader absorption bandwidth while maintaining a small thickness. Consequently, it becomes crucially important to identify the physically realizable upper limit for BW, given a specific $d$ and a desired reflection reduction (hence, a desired absorption level). To address this problem, in Ref. \cite{rozanov2000ultimate} Rozanov exploited the analytical properties of the reflection coefficient function, for linear time-invariant (LTI) passive systems, to establish the theoretical upper bound for the integral of $\ln{| \Gamma(\lambda) |}$ over the entire wavelength spectrum (similar to the well-known Bode-Fano limit for matching networks \cite{pozar2011microwave}) as
\begin{equation}
    \label{rozanov1}
    \left | \int_{0}^{\infty} \ln{| \Gamma(\lambda) |} \, \mathrm{d}\lambda \right | \leq 2 \pi ^2 \mu_s d,
\end{equation}
where $\mu_s$ is the static relative permeability of the absorbing material (in our case $\mu_s = 1$). Incidentally we note that reflection/scattering reduction can be obtained either through absorption or interference effects (e.g., with an anti-reflection coating); the Rozanov bound constrains the former approach and the Bode-Fano limit constrains the latter. Here we define the integral on the left side of Eq. (\ref{rozanov1}) as the Rozanov Integral, denoted by $\mathrm{I}_\mathrm{R}$. To demonstrate the applicability of Eq. (\ref{rozanov1}) for passive LTI systems, in Fig. \ref{fig1}(c) $\mathrm{I}_\mathrm{R}$ has been analytically calculated and plotted for an absorbing material with a lossy Drude-type dispersion for a range of plasma frequencies $\omega_p$. This example shows that as the parameters of the system are varied, in this case increasing $\omega_p$, the Rozanov integral (and therefore the absorbed energy) may increase to a maximum value and then decrease. This maximum value is limited by the right-hand side of Eq. (\ref{rozanov1})—referred to as the Rozanov bound— which represents a theoretical upper limit for $\mathrm{I}_\mathrm{R}$ (see  Fig. \ref{fig1}(c)). Notably, this constraint depends solely on the thickness $d$ in the case of a non-magnetic absorber, preventing any further enhancements in absorption performance for traditional passive LTI absorbers. To further explore the connection between absorption performance and thickness, Eq. (\ref{rozanov1}) can be transformed into the following inequality by assuming a reflection magnitude $\Gamma_0$ within a wavelength range of width BW and full reflection outside \cite{rozanov2000ultimate},
\begin{equation}
    \label{rozanov2}
    \left | \ln{\Gamma_0} \right | \mathrm{BW} \leq 2 \pi ^2 \mu_s d.
\end{equation}
Eq. (\ref{rozanov2}) suggests that lowering the acceptable $\Gamma_0$ (thereby increasing absorption) would lead to a lower BW as expected. As discussed earlier, enhancing absorption while maintaining the same bandwidth requires an increase in the absorber thickness, as illustrated in Figure \ref{fig1}(d). However, many practical applications call for electromagnetic absorbing layers that can provide efficient absorption while maintaining a small thickness and minimal weight, making them suitable for integration into a wide range of devices and systems \cite{ra2015thin}.

A potential approach to break the intrinsic trade-off between bandwidth and thickness, and to explore the parameter region not accessible by passive LTI systems (red shaded regions in Figs. \ref{fig1}(c) and (d)) is to violate one of the fundamental assumptions of the Rozanov bound: the time-invariant nature of the absorber. While this approach has been explored in the literature, the employed temporal modulations were typically limited to `time-switching' \cite{li2021temporal, firestein2022absorption, yang2022broadband}, where one or more material properties undergo instantaneous transitions over time. Although this method holds significant potential for advancing electromagnetic absorbers beyond the limits of LTI systems, it requires precise synchronization of the switching event, ensuring that the pulse is completely contained within the absorber when it occurs. In addition to the difficulty in attaining precise timing, the need to accommodate a pulse with a certain spatial width might impose further constraints on the absorber thickness. Moreover, the reduction of reflections within the designated bandwidth BW often results in a corresponding increase in reflections outside of this BW \cite{tennant1997reflection, chambers2005smart, li2021temporal}. In fact, this is one of the key mechanisms of such time-varying systems, which reduce reflections within a desired BW not only through absorption, but also by effectively redistributing the incident energy beyond the original bandwidth, as also done in phase modulated lossless screens \cite{tennant1997reflection, chambers2005smart}. Finally, we also note that altering material properties instantaneously, or very rapidly, is challenging, especially at optical frequencies \cite{saha2023photonic, hayran2022omega}.

Although time-switching has been the most studied approach for challenging theoretical limits in electromagnetics \cite{hayran2023using} due to its conceptual simplicity, periodic temporal modulations represent an intriguing alternative. Periodically modulated electromagnetic/photonic structures (sometimes referred to as photonic time crystals \cite{saha2023photonic, wang2023metasurface, lyubarov2022amplified, hayran2022omega}) have recently gained significant interest due to the wealth of intriguing physical phenomena they may unveil \cite{yin2022floquet}, but their potential for breaking conventional electromagnetic performance bounds has received less attention \cite{li2019beyond, suwunnarat2019dynamically, mostafa2022coherently}. Similar to time-switching \cite{moussa2023observation, shlivinski2018beyond}, periodically modulating a material can alter the frequency spectrum of the probe wave \cite{zurita2010resonances, koutserimpas2018electromagnetic, galiffi2022photonics}, which suggests that the trade-off between BW and absorption described above may be modified. However, in contrast to time-switching, periodic time modulation can lead to parametric processes \cite{lee2021parametric}, where the probe wave may experience amplification or absorption within the time-modulated medium \cite{sloan2022optical}. While parametric amplification in periodically modulated systems has attracted considerable attention \cite{lyubarov2022amplified, park2022comment, wang2023metasurface, gaxiola2023growing, asadchy2022parametric, koutserimpas2018nonreciprocal, pacheco2023holding}, their absorption aspects have yet to be thoroughly investigated, especially in relation to their implications for overcoming performance limitations in electromagnetics and photonics.

In this study, we explore the properties of periodically time-modulated dissipative and dispersive systems, aiming to enhance the absorption performance of thin absorbers beyond the Rozanov bound. Specifically, we examine a novel approach that combines two distinct absorption mechanisms: material-based and parametric absorption. Our findings reveal that by choosing the proper modulation parameters, the inherent BW-$d$ trade-off of electromagnetic absorbers can be manipulated, and we demonstrate that the parameter space prohibited by the Rozanov bound can be accessed.

\section{Theory: Waves in a Dissipative and Dispersive Time-Modulated Medium} \label{theory}
In this section, we examine how periodic temporal modulations influence the absorption properties of an isotropic homogeneous lossy slab with Drude-type dispersive permittivity, where the electric polarization density $\mathbf{P}$ is related to the electric field $\mathbf{E}$ through
\begin{equation}
    \label{dispersion}
    \frac{\partial^2 \mathbf{P}}{\partial t^2} + \gamma\frac{\partial \mathbf{P}}{\partial t} = \epsilon_0 \omega_\mathrm{p}^2(t) \mathbf{E}.
\end{equation}
Here, $\gamma$ represents the damping coefficient, $\epsilon_0$ denotes the free space permittivity, and the plasma frequency, $\omega_\mathrm{p}(t)$, is subject to a periodic time modulation as
\begin{equation}
    \label{plasma_modulation}
    \omega^2_\mathrm{p}(t) = \omega^2_{\mathrm{p0}}\big(1 + f_{\mathrm{mod}}\sin(\omega_{\mathrm{mod}}t + \varphi_{\mathrm{mod}})\big),
\end{equation}
where  $f_{\mathrm{mod}}$, $\omega_{\mathrm{mod}}$, and $\varphi_{\mathrm{mod}}$ correspond to the modulation amplitude, modulation frequency, and modulation phase, respectively, while $\omega_{\mathrm{p0}}$ is the time-invariant plasma frequency in the absence of any temporal modulation. The plasma frequency, denoted as $\omega_{\mathrm{p}}$, can be dynamically modulated through several strategies. Among these, the most common one involves altering the free carrier density via carrier injection/depletion through electrical gating in electro-optic structures \cite{feigenbaum2010unity} or via optical carrier injection through light absorption \cite{leonard2002ultrafast}. Another approach involves modifying the average effective mass of the electron sea, which may be achieved by redistributing carriers within a non-parabolic band through intraband absorption, induced by an intense optical pump excitation \cite{saha2023photonic}. In the case of metamaterials operating at microwave frequencies, the effective plasma frequency can be altered through electrically controlled varactors, which affect some relevant geometric properties of the unit cells \cite{gil2006tunable}. 

As illustrated in Fig. \ref{fig2}(a), an incident optical pulse, characterized by the spatio-temporal profile presented in Fig. \ref{fig2}(c), interacts with the absorber, while the absorber material undergoes a temporal modulation according to Eq. (\ref{plasma_modulation}). To gain insight into the effect of such a temporal modulation on the incident wave, we start by considering the electromagnetic wave equation in an isotropic homogeneous medium,
\begin{equation}
    \label{wave_eq}
    \frac{\partial^2 \mathbf{D}}{\partial t^2} - \epsilon_0 c_0^2 \nabla^2 \mathbf{E} = 0,
\end{equation}
where $\mathbf{D} = \epsilon_0 \mathbf{E} + \mathbf{P}$ is the electric displacement field. Given a spatially homogeneous time modulation, the conservation of the wavevector $\boldsymbol{k} = k \boldsymbol{r}$ allows us to express the fields as
\begin{equation}
    \label{scalar_fields}
    \mathbf{\Psi} = \textrm{Re}\{\Psi(t) e^{-i k r}\}\boldsymbol{n},
\end{equation}
where $\boldsymbol{n}$ is the polarization vector, $\mathbf{\Psi} = \mathbf{E}, \mathbf{P}, \mathbf{D}$, and $\Psi(t) = \mathrm{E}(t), \mathrm{P}(t), \mathrm{D}(t)$. For our analysis, we assume a highly dissipative material where $\gamma$ is very large compared to the probe frequencies in Eq. (\ref{dispersion}). Consequently, the second order differential term in the left hand side of Eq. (\ref{dispersion}) can be neglected and Eq. (\ref{wave_eq}) can be rewritten as
\begin{equation}
    \label{wave_eq_2}
    \frac{d^2 \mathrm{E}(t)}{d t^2} + \frac{\omega^2_\mathrm{p}(t)}{\gamma} \frac{d \mathrm{E}(t)}{dt} + \left( c_0^2 k^2 + \frac{1}{\gamma} \frac{d \omega^2_\mathrm{p}(t)}{dt}\right) \mathrm{E}(t) = 0.
\end{equation}
Eq. (\ref{wave_eq_2}) bears resemblance to a harmonic oscillator equation where both the resonance frequency and damping coefficient are subject to time modulation. To simplify our analysis, we eliminate the first order differential `damping' term through the following change of variables
\begin{equation}
    \label{variable_change}
    \Theta(t) = e^{\alpha(t)} \mathrm{E}(t), \qquad
    \alpha(t) = \frac{1}{2 \gamma} \int_{0}^{t} \omega^2_\mathrm{p} (t^\prime) \,dt^\prime.
\end{equation}
Eq. (\ref{wave_eq_2}) then becomes
\begin{equation}
    \label{wave_eq_3}
    \frac{d^2 \Theta(t)}{dt^2} + \Omega(t) \Theta(t) = 0,
\end{equation}
where, using Eq. (\ref{plasma_modulation}) and neglecting terms containing frequencies other than $\omega_\mathrm{mod}$ (as they will lead to off-resonance terms that can be neglected to a first order approximation in our subsequent derivation),
\begin{equation}
    \label{transformed_w0}
    \Omega(t) \approx \omega^2_0 \left[ 1 + f_1 \cos(\omega_\mathrm{mod} t + \phi_\mathrm{mod})
    + f_2 \sin(\omega_\mathrm{mod} t + \phi_\mathrm{mod}) \right],
\end{equation}
\begin{equation}
    \label{w0_definition}
    \omega^2_0 = c^2_0 k^2 - \frac{\omega^4_\mathrm{p0}}{4\gamma^2}\left( 1 + \frac{f^2_\mathrm{mod}}{2} \right), \qquad
    f_1 = \frac{\omega^2_\mathrm{p0} \omega_\mathrm{mod}}{2 \gamma \omega^2_0} f_\mathrm{mod}, \qquad
    f_2 = -\frac{\omega^4_\mathrm{p0}}{2 \gamma^2 \omega^2_0} f_\mathrm{mod}.
\end{equation}
For $\omega_\mathrm{mod} \sim 2 \omega_0$, the general solution of Eq. (\ref{wave_eq_3}) can be approximated as \cite{landau2013mechanics}
\begin{equation}
    \label{wave_equation_solution}
    \Theta(t) \approx a_1(t) \cos(\frac{\omega_\mathrm{mod}}{2} t + \frac{\phi_\mathrm{mod}}{2}) + a_2(t) \sin(\frac{\omega_\mathrm{mod}}{2} t + \frac{\phi_\mathrm{mod}}{2}),
\end{equation}
where $a_1(t)$ and $a_2(t)$ are slowly varying functions in time compared to the sinusoidal terms. Inserting Eq. (\ref{wave_equation_solution}) into Eq. (\ref{wave_eq_3}) and neglecting the second order time derivatives of $a_{1,2}(t)$ and any terms containing frequencies other than $\omega_\mathrm{mod} / 2$ (i.e., using a first order approximation \cite{landau2013mechanics}) yields the coupled differential equations
\begin{equation}
    \label{coupled_diff_eq}
    \frac{d}{dt}\mathbf{A} = \mathrm{M}\mathbf{A},
\end{equation}
\begin{equation}
    \label{coupled_diff_eq_variables}
     \mathbf{A} = 
     \begin{bmatrix}
        a_1(t)\\
        a_2(t)
    \end{bmatrix}, \qquad
    \mathrm{M} =
    \frac{\omega^2_0}{2 \omega_\mathrm{mod}}
    \begin{bmatrix}
        f_2 & -f_1 + 2 \Delta\omega \\
        -f_1 - 2\Delta\omega & -f_2
    \end{bmatrix}, \qquad
    \Delta\omega = \frac{4 \omega^2_0 - \omega^2_\mathrm{mod}}{4 \omega^2_0}.
\end{equation}
The solution for $\mathbf{A}$ can be found as
\begin{equation}
    \label{solution_A}
    \mathbf{A} = \kappa_1 \mathbf{V_1} e^{\lambda_1 t} + \kappa_2 \mathbf{V_2} e^{\lambda_2 t},
\end{equation}
where $\kappa_1$ and $\kappa_2$ are constants that are to be determined from the initial conditions, $\mathbf{V_1}$ and $\mathbf{V_2}$ are the eigenvectors of the matrix $\mathrm{M}$, and $\lambda_1$, $\lambda_2$ are the corresponding eigenvalues, which can be found as
\begin{equation}
    \label{eigenvalues}
    \lambda_{1,2} = \pm \frac{\omega^2_0}{2\omega_\mathrm{mod}} \sqrt{f^2_1 + f^2_2 - 4\Delta\omega^2},
\end{equation}
which shows the condition for having exponentially growing and decaying terms in Eq. (\ref{solution_A}), namely, $\Delta\omega^2 < (f^2_1 + f^2_2)/4$. For $\Delta\omega = 0$, the eigenvectors become
\begin{equation}
    \label{eigenvectors}
    \mathbf{V_1} =
    \begin{bmatrix}
        f_2 + \sqrt{f^2_1 + f^2_2}  \\
        -f_1
    \end{bmatrix}, \qquad
    \mathbf{V_2} = 
    \begin{bmatrix}
        f1 \\
        f_2 + \sqrt{f^2_1 + f^2_2} 
    \end{bmatrix}.
\end{equation}
A close inspection of Eqs. (\ref{wave_equation_solution}, \ref{solution_A}-\ref{eigenvectors}) reveals that, for $\Delta\omega = 0$, a purely growing $\Theta(t)$ (i.e., $\kappa_1 \neq 0$, $\kappa_2 = 0$) and a purely decaying $\Theta(t)$ (i.e., $\kappa_1 = 0$, $\kappa_2 \neq 0$)  differ by a phase of $\theta_\mathrm{mod} = \pi$. Furthermore, we note from Eq. (\ref{variable_change}) that the exponential terms for $\mathrm{E}(t)$ become approximately equal to $\textrm{exp}{[(\lambda_{1,2} - \frac{\omega^2_{p0}}{2\gamma}) t]}$. Generally, in periodically modulated non-dispersive non-dissipative materials, one of the eigenfrequencies results in exponentially growing fields, while the other corresponds to exponentially decaying fields \cite{chamanara2018unusual, lee2021parametric}. As the exponentially growing mode tends to dominate, parametric amplification becomes more apparent, whereas observing parametric absorption becomes more challenging \cite{wang2023metasurface}. In our case, however, both exponential terms for $\mathrm{E}(t)$ can be designed to be decaying in time. One way to achieve this is to ensure that $(\omega_\mathrm{p0}^2 / (\gamma \omega_\mathrm{mod}))^2 \ll 1$, so that $f_2$ can be neglected in comparison to $f_1$. In this case, it becomes clear that both exponential terms for $\mathrm{E}(t)$ are exponentially decaying in time as long as $\lvert f_\mathrm{mod} \rvert < 2$ (note that the physically accessible range for $f_\mathrm{mod}$ is $\lvert f_\mathrm{mod} \rvert \le 1$ according to Eq. (\ref{plasma_modulation})). A more general condition to prevent amplification of the probe wave can be found as $\omega_\mathrm{mod} > \omega_\mathrm{p0}^2 / (\sqrt{3} \gamma)$ (together with the assumption made earlier that $\gamma$ is much larger than the probe frequencies, which implies that $\gamma \gg \omega_\mathrm{mod}$). This is further verified through Bloch-Floquet theory calculations, which involve expressing the field quantities and time-varying parameters in terms of Bloch-Floquet expansions and solving for the complex eigenfrequency at each wavevector by using the orthogonality of the frequency harmonics \cite{chamanara2018unusual, taravati2020space}. The result of this analysis is given in Fig. \ref{fig2}(b), which shows the electromagnetic bandstructure resulting from the dissipative and dispersive periodic temporal modulation. The bandstructure is characterized by a momentum bandgap, in which the imaginary part of the eigenfrequency bifurcates into two separate values, both of which are indeed negative, as independently verified through the theory outlined here and the standard Bloch-Floquet method.

In a homogeneous time-varying medium of this type, when a propagating wave has a wavevector falling within the bandgap, both modes will generally be excited (i.e., both $\kappa_1$ and $\kappa_2$ will become non-zero), one with higher and the other with lower decay rate relative to the time-invariant scenario. Consequently, the overall absorption performance will not be significantly affected compared to the time-invariant case. While adjusting the modulation phase in relation to a propagating probe wave can selectively excite the specific eigenmode that favors higher absorption, consistent with theoretical predictions, this generally necessitates a traveling wave-type spatio-temporal modulation \cite{galiffi2019broadband}. In contrast, here, the absorbing slab backed by a reflector generates a standing probe wave within the slab, allowing the wave to oscillate either in phase (increased absorption, when $\kappa_1 = 0$ and $\kappa_2 \neq 0$) or out of phase (reduced absorption, when $\kappa_1 \neq 0$ and $\kappa_2 = 0$) with the temporal modulation. Thus, by carefully selecting the modulation parameters, it may be possible to enhance the absorption performance effectively, potentially surpassing the limits of LTI absorbers.

\section{Methods and Results: Going Beyond the Rozanov Bound} \label{methods}

\begin{figure*}
 \centering
  \includegraphics[ width=0.8\linewidth, keepaspectratio]{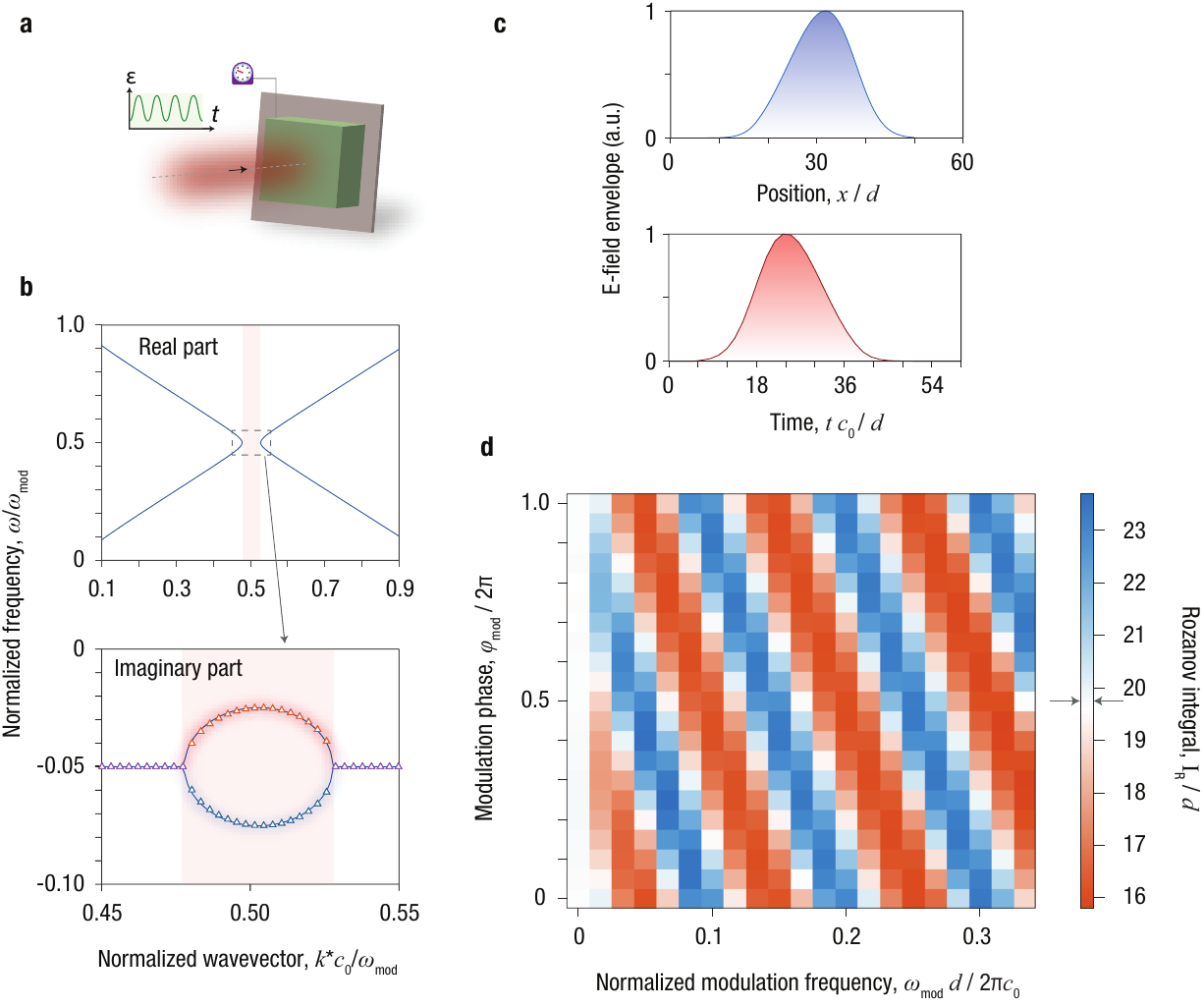}
\caption{(a) Schematic diagram illustrating an incident pulse interacting with an electromagnetic absorber with periodically modulated permittivity. (b) Bandstructure of the time-modulated material, calculated through the Floquet theorem. The emergence of a momentum bandgap in the bandstructure due to the periodic temporal modulation (upper panel) results in the imaginary part of the eigenfrequency splitting into two distinct values within the gap (lower panel). Triangular markers denote results obtained using the theory outlined in Section \ref{theory}. Material and modulation parameters are as follows: $\omega_{\mathrm{p0}} = 10 \omega_{\mathrm{mod}}$, $\gamma = 1000 \omega_{\mathrm{mod}}$, and $f_\mathrm{mod} = 1$. (c) Spatio-temporal characteristics of the considered incoming pulse. (d) Rozanov integral values, $\mathrm{I}_\mathrm{R}$, for a range of modulation parameters. Absorber properties are as follows: $\omega_{\mathrm{p0}} d / 2\pi c_0 = 6$, $\gamma d / 2\pi c_0 = 330$, and $f_{\mathrm{mod}} = 0.4$. Colorbar arrows in (c) indicate the Rozanov limit.
}
    \label{fig2}
\end{figure*}

Following the theoretical insight outlined in the previous section, in the following we investigate the reflection properties of periodically time-modulated absorbers and compare them against the Rozanov bound. It is important to note that, in the context of time-varying systems, where frequency is not conserved, the reflection coefficient should not generally be represented by $\Gamma(\omega)$, but instead by a function $\Gamma(\omega, \omega')$ that relates the input and output electric fields as $\mathbf{E_\mathrm{out}(\omega)} = \int \Gamma(\omega,\omega') \mathbf{E_\mathrm{in}(\omega')} \,d\omega'$, as done for other time-varying response functions \cite{sloan2022optical} Nevertheless, for a specific incoming signal, one may also define an equivalent input-dependent reflection coefficient $\Gamma_{\mathrm{eq}}(\omega)$ via the relation $\mathbf{E}_\mathrm{out}(\omega) = \Gamma_{\mathrm{eq}}(\omega) \mathbf{E}_\mathrm{in}(\omega)$  to compute the integral $\mathrm{I}_\mathrm{R}$ within the frequency range of the incident pulse (and assuming total reflection outside this range). 
Note that, due to the frequency non-conserving nature of the time-modulated system, this equivalent reflection coefficient will only be valid for the considered input field with its specific temporal/spectral profile, and will only be useful for the case of sufficiently broadband signals for which the generated reflected fields at frequencies beyond the bandwidth of the original signal would be negligible. The case of narrow-band incident pulses will be treated differently in the following, demonstrating however similar results. Thus, we consider here an incoming pulse with an ultra-wide BW, corresponding to a normalized wavelength range from $\lambda / d = 0.02$ to $\lambda / d = 1000$, and we numerically compute the integral $\mathrm{I}_\mathrm{R}$ over this broad wavelength range to evaluate the absorption performance of the system. It should also be noted that for a generic time-varying system the magnitude of the reflection coefficient may be larger than unity at certain wavelengths, due to parametric amplification and/or harmonic generation, which would imply a reduction of the Rozanov integral $\mathrm{I}_\mathrm{R}$, corresponding to a degradation, not an enhancement, of the absorption performance over the entire spectrum. In contrast, our approach based on parametrically-enhanced absorption is expected to increase $\mathrm{I}_\mathrm{R}$, potentially beyond the Rozanov bound (the right-hand-side of Eq. (\ref{rozanov1})). To demonstrate this opportunity for enhanced absorption performance, the Rozanov integral for the input given in Fig. \ref{fig2}(c) has been numerically calculated using the finite-difference time-domain method \cite{solutions2022lumerical} for a range of modulation parameters, as shown in Fig. \ref{fig2}(d). It is evident that, for $\omega_\mathrm{mod} = 0$ (i.e., time-invariant case), the integral $\mathrm{I}_\mathrm{R}$ consistently remains below the Rozanov bound. However, when the modulation frequency $\omega_\mathrm{mod}$ takes on non-zero values, the Rozanov integral can exceed the Rozanov bound  for certain modulation parameters. It is important to note that the frequency spectrum of the input field is sufficiently broad, allowing each of the considered $\omega_{mod}$ to enable the parametric processes that enhance absorption (which occurs when the modulation frequency is twice the signal frequency). Moreover, as expected from our theoretical analysis, the absorption performance depends on the modulation phase $\varphi_{mod}$ indicating an optimal $\varphi_{mod}$ for each modulation frequency, for which the eigenmode with higher absorption can be selectively excited. It is also worth mentioning that the spatial width of the impinging pulse (see Fig. \ref{fig2}(c)) is significantly larger than the thickness ($d$) of the absorber. This situation would not be suitable for time-switched systems \cite{li2021temporal, firestein2022absorption}, where the switching needs to take place when the pulse is entirely contained within the system.

\begin{figure*}
 \centering
  \includegraphics[ width=0.8\linewidth, keepaspectratio]{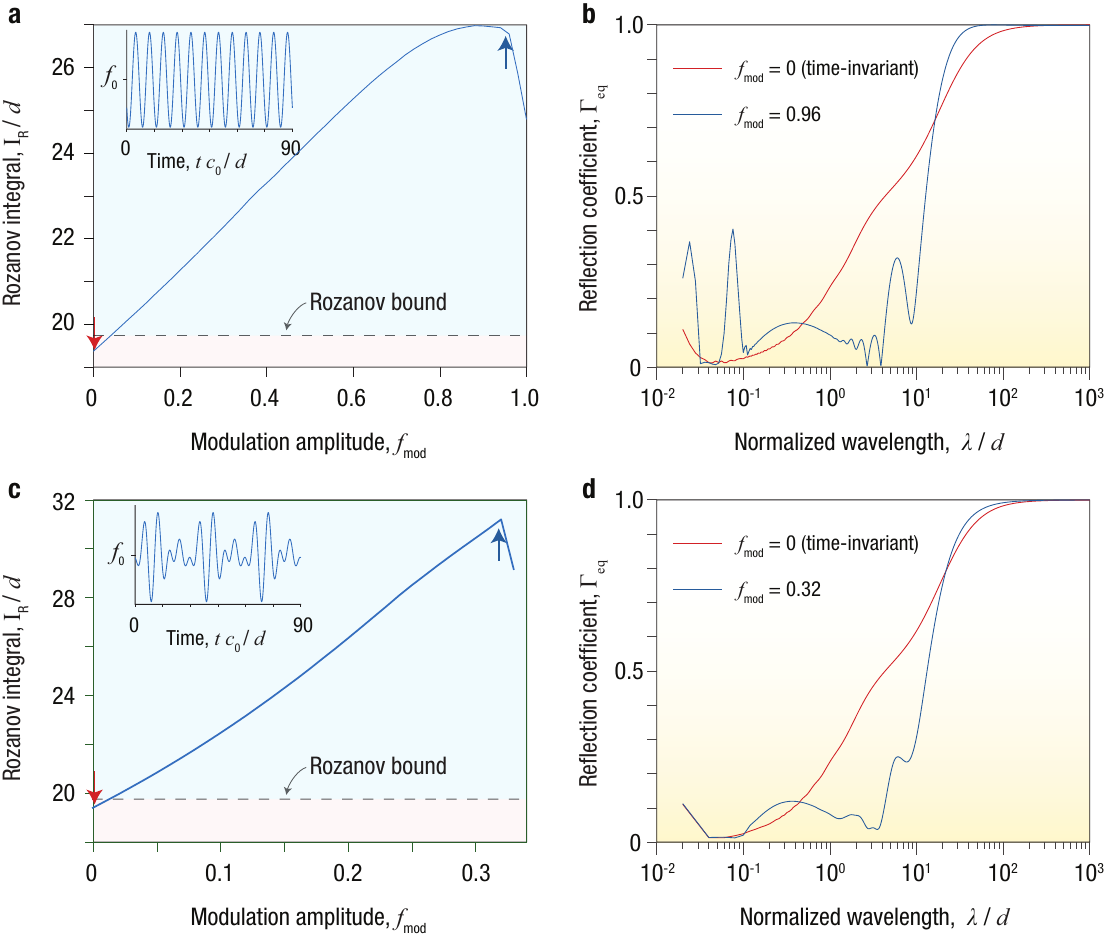}
\caption{(a) Computed Rozanov integral $\mathrm{I}_\mathrm{R}$ as a function of the modulation amplitude. (b) Superimposed reflection coefficient spectra for $f_\mathrm{mod}$ = 0 (time-invariant), and $f_\mathrm{mod}$ = 0.96 (denoted with red and blue arrows in (a), respectively). 
Absorber properties are the same as in Fig. \ref{fig2}, with $\varphi_\mathrm{mod}/\pi = 0.60$ and $\omega_\mathrm{mod} d / 2\pi c_0 = 0.133$. (c,d) same as (a) and (b), but with a multi-harmonic modulation function with modulation frequencies $\omega_{\mathrm{mod}_{1,2,3}} d / 2\pi c_0 = 
\{0.100, 0.133, 0.166\}$, modulation phases $\varphi_{\mathrm{mod}_{1,2,3}} / 2\pi = 
\{0.90, 0.60, 0.25\}$, and the same modulation amplitudes.
}
    \label{fig3}
\end{figure*}

To further investigate the effect of the temporal modulation on the absorption characteristics, Fig. \ref{fig3}(a) shows the calculated $\mathrm{I}_\mathrm{R}$ for the full range of modulation amplitudes $f_\mathrm{mod}$, for the same broadband incident pulse. Remarkably, Fig. \ref{fig3}(a) shows that, in this scenario, even for small modulation amplitudes ($f_\mathrm{mod} \sim 0.05$), it is possible to surpass the Rozanov bound. Interestingly, the $\mathrm{I}_\mathrm{R}$-$f_\mathrm{mod}$ curve presents a peak, after which an increase in $f_\mathrm{mod}$ leads to a decrease in the integral value of $\mathrm{I}_\mathrm{R}$. We ascribe this phenomenon to the intricate interplay of harmonic redistribution effects and parametric processes in this time-varying medium. Although parametric processes could reduce reflection at specific wavelengths, the harmonic redistribution due to the temporal modulation may enhance the reflection coefficient at adjacent wavelengths. Consequently, this may establish an optimal point where the integral $\mathrm{I}_\mathrm{R}$ reaches its maximum value. To examine more closely the absorption characteristics of the time-modulated system, Fig. \ref{fig3}(b) plots the superimposed reflection coefficient for the time-invariant ($f_\mathrm{mod} = 0$) and time-modulated ($f_\mathrm{mod} = 0.96$) case. It is clear that, when the absorber is temporally modulated, the reflection diminishes considerably across an extensive wavelength spectrum, ranging from $\lambda / d = 0.5$ to $\lambda / d = 15$, while the reflection experiences an increase at wavelengths adjacent to this range. Overall, the integral $\mathrm{I}_\mathrm{R}$ increases beyond the Rozanov bound, given that it is primarily influenced by the wavelength range where the reflection experiences a strong reduction. 

Moreover, to further improve the absorption performance for a broadband impinging pulse, the modulation frequency can be made multi-harmonic, i.e,
\begin{equation}
    \label{plasma_modulation_multiple}
    \omega^2_\mathrm{p}(t) = \omega^2_{\mathrm{p0}}\big(1 + f_{\mathrm{mod}} \sum_n \sin(\omega_{\mathrm{mod}_n}t + \varphi_{\mathrm{mod}_n})\big),
\end{equation}
which allows optimizing absorption for multiple frequency ranges within the signal bandwidth. As demonstrated in Figs. \ref{fig3}(c) and (d), this method has the potential advantage of requiring smaller modulation amplitudes for each modulation frequency component to achieve absorption performance comparable to the case of a single-frequency modulation, owing to the possibility of combining reflection reductions across multiple frequency ranges. It is also worth noting that the reflections occurring outside the enhanced absorption window do not experience a significant increase, as is the case with a single-frequency modulation (see Figs. \ref{fig3}(b) and (d)). This is mainly because the lower modulation amplitudes are not sufficient to induce strong harmonic generations at these wavelengths.

\begin{figure*}
 \centering
  \includegraphics[ width=0.8\linewidth, keepaspectratio]{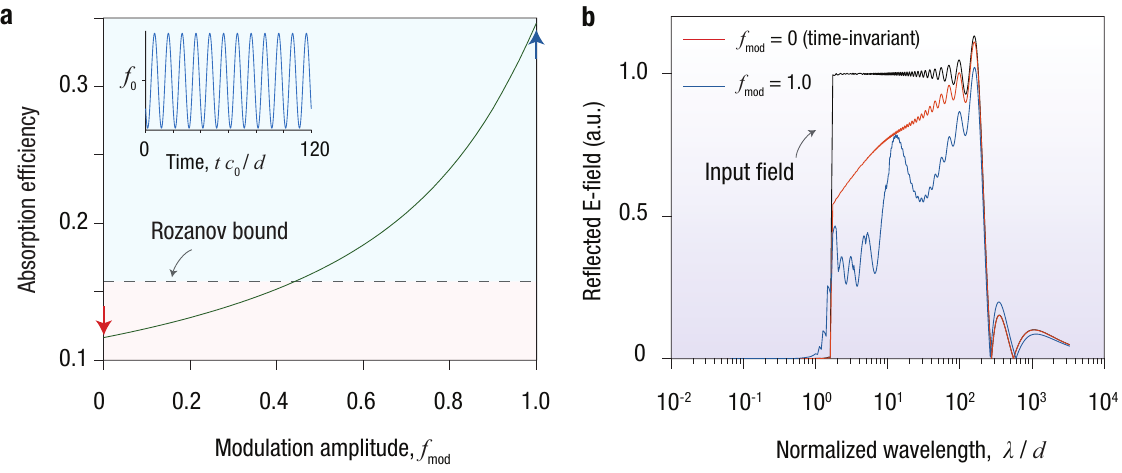}
\caption{(a) Computed absorption efficiency as a function of the modulation amplitude. (b) Superimposed reflected electric field spectra for $f_\mathrm{mod}$ = 0 (time-invariant) and $f_\mathrm{mod}$ = 1.0 (denoted with red and blue arrows in (a), respectively). Absorber properties are as follows: 
$\omega_\mathrm{p0} d / 2\pi c_0 = 5.5$, $\gamma d / 2\pi c_0 = 100$, $\varphi_\mathrm{mod}/\pi = 0.45$, and $\omega_\mathrm{mod} d / 2\pi c_0 = 0.1$.
}
    \label{fig4}
\end{figure*}

While the previous analysis focused on an ultrabroadband input field, exploring the time-varying absorber response for a field with relatively narrow bandwidth would provide further insight. In this situation, the integral $\mathrm{I}_\mathrm{R}$ may not serve as a reliable measure of the absorption performance, as the equivalent reflection coefficient, defined as above, would take extremely large values or even become infinite beyond the input pulse bandwidth. Thus, following a similar approach as in Ref. \cite{firestein2022absorption}, a figure of merit denoted as absorption efficiency (AE) is defined to evaluate the absorption performance, and compare it against the Rozanov bound, over the entire electromagnetic spectrum:
\begin{equation}
    \label{absorption_rate}
    \mathrm{AE} = 1 - \frac
    {\int_{-\infty}^{\infty} \! \left | \mathrm{E}_{\mathrm{ref}} (\lambda) \right |^2 \, \mathrm{d}\lambda}
    {\int_{-\infty}^{\infty} \! \left | \mathrm{E}_{\mathrm{inc}} (\lambda) \right |^2 \, \mathrm{d}\lambda},
\end{equation}
where $\mathrm{E}_{\mathrm{ref}}$ and $\mathrm{E}_{\mathrm{inc}}$ are respectively the reflected and incident electric field spectrum. Note that this formulation differs from that in Ref. \cite{firestein2022absorption}, where the integration is done over frequency rather than wavelength. This choice is to ensure a direct comparison with the Rozanov bound, which is derived by performing an integration in the complex wavelength domain. According to this FOM definition, the corresponding Rozanov bound for the LTI case can be written as
\begin{equation}
    \label{absorption_rate_rozanov}
    \mathrm{AE}_{\mathrm{R}} = 1 - \frac
    {\int_{-\infty}^{\infty} \! \left | \Gamma_0 \mathrm{E}_{\mathrm{inc}} (\lambda) \right |^2 \, \mathrm{d}\lambda}
    {\int_{-\infty}^{\infty} \! \left | \mathrm{E}_{\mathrm{inc}} (\lambda) \right |^2 \, \mathrm{d}\lambda},
\end{equation}
where $\Gamma_0$ assumes the lowest value it is allowed to take for the given bandwidth BW, according to Eq. (\ref{rozanov2}), i.e.,
\begin{equation}
    \label{absorption_rate_rozanov2}
    \Gamma_0 = e^{{- 2 \pi^2 d}/\mathrm{BW}}
\end{equation}
within the considered wavelength range, and $\Gamma_0 = 1$ outside this range. Note that for an accurate comparison with the Rozanov bound, the incident pulse should exhibit a sufficiently well-defined BW, while having minimal energy outside this range, as in the case in Fig. \ref{fig4}. As an example, the absorption efficiency for a signal with a wavelength bandwidth ranging from $\lambda / d = 1.66$ to $\lambda / d = 200$ was calculated for a range of modulation parameters (see Fig. \ref{fig4}(a)). Similar to the trend observed in Fig. \ref{fig3}(a), these results clearly show that the absorption performance improves in the time-varying case and can surpass the Rozanov bound by a significant amount. To further clarify the reflection response of the time-modulated absorber, Figure \ref{fig4}(b) displays the superimposed reflection spectrum for the time-invariant and time-varying cases: When the material is temporally modulated, the reflected field is indeed reduced almost everywhere within the incident pulse bandwidth. Although extra harmonics are generated at shorter wavelengths beyond this bandwidth (as can be observed in Fig. \ref{fig4}(b)), the strong reduction of in-band reflection results in an overall increase of absorption efficiency that allows surpassing the Rozanov bound.

\section{Summary} \label{summary}
In conclusion, we have examined the relatively unexplored absorption aspects of periodically time-modulated lossy materials, focusing on their implications to surpass the electromagnetic absorption limits of linear time-invariant systems. Our findings -- in terms of two different figures of merit (Rozanov integral of the reflection coefficient, and absorption efficiency) -- consistently show that by selecting appropriate modulation parameters, it is possible to drastically enhance the absorption performance of the system compared to its time-invariant counterpart and even surpass the Rozanov bound for electromagnetic absorption, fundamentally breaking the tradeoff between thickness, bandwidth, and reflection reduction. In contrast to time-switched mechanisms, our method does not require exact timing to guarantee that the pulse is entirely contained within the absorber; instead, the considered periodic modulation is continuously applied as the pulse enters the absorbing medium. As a result, our platform inherently addresses the issue of impedance mismatch at the entrance interface between the absorber and free-space, whereas in time-switched platforms the impedance mismatch problem is typically resolved through additional optimizations. Moreover, our approach ensures that the absorber thickness is independent of the pulse spatial width, whereas in time-switched platforms the minimum slab thickness is typically limited by the pulse width. It is also worth mentioning that since both eigenmodes in the momentum bandgap of the periodically modulated structure have a decaying nature, due to the high intrinsic losses of the considered material, the system is free from any instabilities, namely, unbounded temporal oscillations. Ultimately, we hope that our work may provide a new route for the design of advanced electromagnetic absorbers that can surpass conventional performance bounds.

\begin{backmatter}
    \bmsection{Funding} Air Force Office of Scientific Research (FA9550-22-1-0204); Office of Naval Research (N00014-22-1-2486).
    
    
    
\end{backmatter}

\bibliography{beyond_rozanov_periodic.bbl}

\begin{thebibliography}{10}
\newcommand{\enquote}[1]{``#1''}

\bibitem{salisbury1952absorbent}
W.~W. Salisbury, \enquote{Absorbent body for electromagnetic waves,}  (1952).
  US Patent 2,599,944.

\bibitem{knott1980performance}
E.~Knott and K.~Langseth, \enquote{Performance degradation of jaumann absorbers
  due to curvature,} {\protect\JournalTitle{IEEE Transactions on Antennas and
  Propagation}} \textbf{28}, 137--139 (1980).

\bibitem{dallenbach1938reflection}
W.~Dallenbach and W.~Kleinsteuber, \enquote{Reflection and absorption of
  decimeter-waves by plane dielectric layers,} {\protect\JournalTitle{Hochfreq.
  u Elektroak}} \textbf{51}, 152--156 (1938).

\bibitem{amin2013ultra}
M.~Amin, M.~Farhat, and H.~Ba{\u{g}}c{\i}, \enquote{An ultra-broadband
  multilayered graphene absorber,} {\protect\JournalTitle{Optics Express}}
  \textbf{21}, 29938--29948 (2013).

\bibitem{saib2006carbon}
A.~Saib, L.~Bednarz, R.~Daussin, C.~Bailly, X.~Lou, J.-M. Thomassin,
  C.~Pagnoulle, C.~Detrembleur, R.~J{\'e}r{\^o}me, and I.~Huynen,
  \enquote{Carbon nanotube composites for broadband microwave absorbing
  materials,} {\protect\JournalTitle{IEEE Transactions on Microwave Theory and
  Techniques}} \textbf{54}, 2745--2754 (2006).

\bibitem{dayal2013design}
G.~Dayal and S.~A. Ramakrishna, \enquote{Design of multi-band metamaterial
  perfect absorbers with stacked metal--dielectric disks,}
  {\protect\JournalTitle{Journal of Optics}} \textbf{15}, 055106 (2013).

\bibitem{sun2012broadband}
L.~Sun, H.~Cheng, Y.~Zhou, and J.~Wang, \enquote{Broadband metamaterial
  absorber based on coupling resistive frequency selective surface,}
  {\protect\JournalTitle{Optics Express}} \textbf{20}, 4675--4680 (2012).

\bibitem{wakatsuchi2013waveform}
H.~Wakatsuchi, S.~Kim, J.~J. Rushton, and D.~F. Sievenpiper,
  \enquote{Waveform-dependent absorbing metasurfaces,}
  {\protect\JournalTitle{Physical Review Letters}} \textbf{111}, 245501 (2013).

\bibitem{nam2014solar}
Y.~Nam, Y.~X. Yeng, A.~Lenert, P.~Bermel, I.~Celanovic, M.~Solja{\v{c}}i{\'c},
  and E.~N. Wang, \enquote{Solar thermophotovoltaic energy conversion systems
  with two-dimensional tantalum photonic crystal absorbers and emitters,}
  {\protect\JournalTitle{Solar Energy Materials and Solar Cells}} \textbf{122},
  287--296 (2014).

\bibitem{michielssen1993design}
E.~Michielssen, J.-M. Sajer, S.~Ranjithan, and R.~Mittra, \enquote{Design of
  lightweight, broad-band microwave absorbers using genetic algorithms,}
  {\protect\JournalTitle{IEEE Transactions on Microwave Theory and Techniques}}
  \textbf{41}, 1024--1031 (1993).

\bibitem{cui2005application}
S.~Cui and D.~S. Weile, \enquote{Application of a parallel particle swarm
  optimization scheme to the design of electromagnetic absorbers,}
  {\protect\JournalTitle{IEEE Transactions on Antennas and Propagation}}
  \textbf{53}, 3616--3624 (2005).

\bibitem{hou2020customized}
J.~Hou, H.~Lin, W.~Xu, Y.~Tian, Y.~Wang, X.~Shi, F.~Deng, and L.~Chen,
  \enquote{Customized inverse design of metamaterial absorber based on
  target-driven deep learning method,} {\protect\JournalTitle{IEEE Access}}
  \textbf{8}, 211849--211859 (2020).

\bibitem{shrekenhamer2013liquid}
D.~Shrekenhamer, W.-C. Chen, and W.~J. Padilla, \enquote{Liquid crystal tunable
  metamaterial absorber,} {\protect\JournalTitle{Physical Review Letters}}
  \textbf{110}, 177403 (2013).

\bibitem{lei2016magnetically}
M.~Lei, N.~Feng, Q.~Wang, Y.~Hao, S.~Huang, and K.~Bi, \enquote{Magnetically
  tunable metamaterial perfect absorber,} {\protect\JournalTitle{Journal of
  Applied Physics}} \textbf{119}, 244504 (2016).

\bibitem{kats2012ultra}
M.~A. Kats, D.~Sharma, J.~Lin, P.~Genevet, R.~Blanchard, Z.~Yang, M.~M.
  Qazilbash, D.~Basov, S.~Ramanathan, and F.~Capasso, \enquote{Ultra-thin
  perfect absorber employing a tunable phase change material,}
  {\protect\JournalTitle{Applied Physics Letters}} \textbf{101}, 221101 (2012).

\bibitem{zeng2020electromagnetic}
X.~Zeng, X.~Cheng, R.~Yu, and G.~D. Stucky, \enquote{Electromagnetic microwave
  absorption theory and recent achievements in microwave absorbers,}
  {\protect\JournalTitle{Carbon}} \textbf{168}, 606--623 (2020).

\bibitem{rozanov2000ultimate}
K.~N. Rozanov, \enquote{Ultimate thickness to bandwidth ratio of radar
  absorbers,} {\protect\JournalTitle{IEEE Transactions on Antennas and
  Propagation}} \textbf{48}, 1230--1234 (2000).

\bibitem{pozar2011microwave}
D.~M. Pozar, \emph{Microwave Engineering} (John Wiley \& Sons, 2011).

\bibitem{ra2015thin}
Y.~Ra’di, C.~R. Simovski, and S.~A. Tretyakov, \enquote{Thin perfect
  absorbers for electromagnetic waves: theory, design, and realizations,}
  {\protect\JournalTitle{Physical Review Applied}} \textbf{3}, 037001 (2015).

\bibitem{li2021temporal}
H.~Li and A.~Al{\`u}, \enquote{Temporal switching to extend the bandwidth of
  thin absorbers,} {\protect\JournalTitle{Optica}} \textbf{8}, 24--29 (2021).

\bibitem{firestein2022absorption}
C.~Firestein, A.~Shlivinski, and Y.~Hadad, \enquote{Absorption and scattering
  by a temporally switched lossy layer: Going beyond the rozanov bound,}
  {\protect\JournalTitle{Physical Review Applied}} \textbf{17}, 014017 (2022).

\bibitem{yang2022broadband}
X.~Yang, E.~Wen, and D.~F. Sievenpiper, \enquote{Broadband time-modulated
  absorber beyond the {Bode-Fano} limit for short pulses by energy trapping,}
  {\protect\JournalTitle{Physical Review Applied}} \textbf{17}, 044003 (2022).

\bibitem{tennant1997reflection}
A.~Tennant, \enquote{Reflection properties of a phase modulating planar
  screen,} {\protect\JournalTitle{Electronics Letters}} \textbf{33}, 1768--1769
  (1997).

\bibitem{chambers2005smart}
B.~Chambers and A.~Tennant, \enquote{A smart radar absorber based on the
  phase-switched screen,} {\protect\JournalTitle{IEEE Transactions on Antennas
  and Propagation}} \textbf{53}, 394--403 (2005).

\bibitem{saha2023photonic}
S.~Saha, O.~Segal, C.~Fruhling, E.~Lustig, M.~Segev, A.~Boltasseva, and V.~M.
  Shalaev, \enquote{Photonic time crystals: a materials perspective,}
  {\protect\JournalTitle{Optics Express}} \textbf{31}, 8267--8273 (2023).

\bibitem{hayran2022omega}
Z.~Hayran, J.~B. Khurgin, and F.~Monticone, \enquote{$\hbar$$\omega$ versus
  $\hbar$k: {Dispersion} and energy constraints on time-varying photonic
  materials and time crystals,} {\protect\JournalTitle{Optical Materials
  Express}} \textbf{12}, 3904--3917 (2022).

\bibitem{hayran2023using}
Z.~Hayran and F.~Monticone, \enquote{Using time-varying systems to challenge
  fundamental limitations in electromagnetics: Overview and summary of
  applications.} {\protect\JournalTitle{IEEE Antennas and Propagation
  Magazine}}  (2023).

\bibitem{wang2023metasurface}
X.~Wang, M.~S. Mirmoosa, V.~S. Asadchy, C.~Rockstuhl, S.~Fan, and S.~A.
  Tretyakov, \enquote{Metasurface-based realization of photonic time crystals,}
  {\protect\JournalTitle{Science Advances}} \textbf{9}, eadg7541 (2023).

\bibitem{lyubarov2022amplified}
M.~Lyubarov, Y.~Lumer, A.~Dikopoltsev, E.~Lustig, Y.~Sharabi, and M.~Segev,
  \enquote{Amplified emission and lasing in photonic time crystals,}
  {\protect\JournalTitle{Science}} \textbf{377}, 425--428 (2022).

\bibitem{yin2022floquet}
S.~Yin, E.~Galiffi, and A.~Al{\`u}, \enquote{Floquet metamaterials,}
  {\protect\JournalTitle{ELight}} \textbf{2}, 1--13 (2022).

\bibitem{li2019beyond}
H.~Li, A.~Mekawy, and A.~Al{\`u}, \enquote{Beyond {Chu}’s limit with floquet
  impedance matching,} {\protect\JournalTitle{Physical Review Letters}}
  \textbf{123}, 164102 (2019).

\bibitem{suwunnarat2019dynamically}
S.~Suwunnarat, D.~Halpern, H.~Li, B.~Shapiro, and T.~Kottos,
  \enquote{Dynamically modulated perfect absorbers,}
  {\protect\JournalTitle{Physical Review A}} \textbf{99}, 013834 (2019).

\bibitem{mostafa2022coherently}
M.~Mostafa, A.~D{\'\i}az-Rubio, M.~Mirmoosa, and S.~Tretyakov,
  \enquote{Coherently time-varying metasurfaces,}
  {\protect\JournalTitle{Physical Review Applied}} \textbf{17}, 064048 (2022).

\bibitem{moussa2023observation}
H.~Moussa, G.~Xu, S.~Yin, E.~Galiffi, Y.~Ra’di, and A.~Al{\`u},
  \enquote{Observation of temporal reflection and broadband frequency
  translation at photonic time interfaces,} {\protect\JournalTitle{Nature
  Physics}} pp. 1--6 (2023).

\bibitem{shlivinski2018beyond}
A.~Shlivinski and Y.~Hadad, \enquote{Beyond the {Bode-Fano} bound: {Wideband}
  impedance matching for short pulses using temporal switching of
  transmission-line parameters,} {\protect\JournalTitle{Physical Review
  Letters}} \textbf{121}, 204301 (2018).

\bibitem{zurita2010resonances}
J.~R. Zurita-S{\'a}nchez and P.~Halevi, \enquote{Resonances in the optical
  response of a slab with time-periodic dielectric function $\varepsilon$ (t),}
  {\protect\JournalTitle{Physical Review A}} \textbf{81}, 053834 (2010).

\bibitem{koutserimpas2018electromagnetic}
T.~T. Koutserimpas and R.~Fleury, \enquote{Electromagnetic waves in a time
  periodic medium with step-varying refractive index,}
  {\protect\JournalTitle{IEEE Transactions on Antennas and Propagation}}
  \textbf{66}, 5300--5307 (2018).

\bibitem{galiffi2022photonics}
E.~Galiffi, R.~Tirole, S.~Yin, H.~Li, S.~Vezzoli, P.~A. Huidobro, M.~G.
  Silveirinha, R.~Sapienza, A.~Al{\`u}, and J.~Pendry, \enquote{Photonics of
  time-varying media,} {\protect\JournalTitle{Advanced Photonics}} \textbf{4},
  014002--014002 (2022).

\bibitem{lee2021parametric}
S.~Lee, J.~Park, H.~Cho, Y.~Wang, B.~Kim, C.~Daraio, and B.~Min,
  \enquote{Parametric oscillation of electromagnetic waves in momentum band
  gaps of a spatiotemporal crystal,} {\protect\JournalTitle{Photonics
  Research}} \textbf{9}, 142--150 (2021).

\bibitem{sloan2022optical}
J.~Sloan, N.~Rivera, J.~D. Joannopoulos, and M.~Solja{\v{c}}i{\'c},
  \enquote{Optical properties of dispersive time-dependent materials,}
  {\protect\JournalTitle{arXiv preprint arXiv:2211.16166}}  (2022).

\bibitem{park2022comment}
J.~Park, H.~C. Park, K.~Lee, J.~Shin, J.-W. Ryu, W.~Jeon, N.~Park, and B.~Min,
  \enquote{Comment on" amplified emission and lasing in photonic time
  crystals",} {\protect\JournalTitle{arXiv preprint arXiv:2211.14832}}  (2022).

\bibitem{gaxiola2023growing}
J.~Gaxiola-Luna and P.~Halevi, \enquote{Growing fields in a temporal photonic
  (time) crystal with a square profile of the permittivity $\varepsilon$ (t),}
  {\protect\JournalTitle{Applied Physics Letters}} \textbf{122}, 011702 (2023).

\bibitem{asadchy2022parametric}
V.~Asadchy, A.~Lamprianidis, G.~Ptitcyn, M.~Albooyeh, T.~Karamanos, R.~Alaee,
  S.~Tretyakov, C.~Rockstuhl, S.~Fan \emph{et~al.}, \enquote{Parametric mie
  resonances and directional amplification in time-modulated scatterers,}
  {\protect\JournalTitle{Physical Review Applied}} \textbf{18}, 054065 (2022).

\bibitem{koutserimpas2018nonreciprocal}
T.~T. Koutserimpas and R.~Fleury, \enquote{Nonreciprocal gain in non-hermitian
  time-floquet systems,} {\protect\JournalTitle{Physical Review Letters}}
  \textbf{120}, 087401 (2018).

\bibitem{pacheco2023holding}
V.~Pacheco-Pe{\~n}a, Y.~Kiasat, D.~M. Sol{\'\i}s, B.~Edwards, and N.~Engheta,
  \enquote{Holding and amplifying electromagnetic waves with temporal
  non-foster metastructures,} {\protect\JournalTitle{arXiv preprint
  arXiv:2304.03861}}  (2023).

\bibitem{feigenbaum2010unity}
E.~Feigenbaum, K.~Diest, and H.~A. Atwater, \enquote{Unity-order index change
  in transparent conducting oxides at visible frequencies,}
  {\protect\JournalTitle{Nano Letters}} \textbf{10}, 2111--2116 (2010).

\bibitem{leonard2002ultrafast}
S.~Leonard, H.~Van~Driel, J.~Schilling, and R.~Wehrspohn, \enquote{Ultrafast
  band-edge tuning of a two-dimensional silicon photonic crystal via
  free-carrier injection,} {\protect\JournalTitle{Physical Review B}}
  \textbf{66}, 161102 (2002).

\bibitem{gil2006tunable}
I.~Gil, J.~Bonache, J.~Garcia-Garcia, and F.~Martin, \enquote{Tunable
  metamaterial transmission lines based on varactor-loaded split-ring
  resonators,} {\protect\JournalTitle{IEEE Transactions on Microwave Theory and
  Techniques}} \textbf{54}, 2665--2674 (2006).

\bibitem{landau2013mechanics}
L.~D. Landau and E.~M. Lifshitz, \emph{Mechanics: Course of Theoretical
  Physics}, vol.~1 (Butterworth-Heinemann, 1976). \S 27. Parametric resonance,
  p. 80-83.

\bibitem{chamanara2018unusual}
N.~Chamanara, Z.-L. Deck-L{\'e}ger, C.~Caloz, and D.~Kalluri, \enquote{Unusual
  electromagnetic modes in space-time-modulated dispersion-engineered media,}
  {\protect\JournalTitle{Physical Review A}} \textbf{97}, 063829 (2018).

\bibitem{taravati2020space}
S.~Taravati and A.~A. Kishk, \enquote{Space-time modulation: {Principles} and
  applications,} {\protect\JournalTitle{IEEE Microwave Magazine}} \textbf{21},
  30--56 (2020).

\bibitem{galiffi2019broadband}
E.~Galiffi, P.~Huidobro, and J.~B. Pendry, \enquote{Broadband nonreciprocal
  amplification in luminal metamaterials,} {\protect\JournalTitle{Physical
  Review Letters}} \textbf{123}, 206101 (2019).

\bibitem{solutions2022lumerical}
{\relax Ansys Lumerical FDTD Solutions 2022}.
  Https://www.lumerical.com/products/fdtd/.

\end{thebibliography}

\end{document}